\documentclass{ws-ijmpb}

\begin{document}

\markboth{S.Aubry and G.Kopidakis}
{Ultrafast Catalytic Transfer of 
   Electrons}

\catchline{}{}{}

\title{A Nonlinear Dynamical Model for
   Ultrafast Catalytic Transfer of 
   Electrons at Zero Temperature}

\author{S. Aubry \footnote{email: \texttt{aubry@llb.saclay.cea.fr}}}

\address{Laboratoire 
L\'eon Brillouin (CEA-CNRS), CEA Saclay\\
91191-Gif-sur-Yvette Cedex, France}

\author{G. Kopidakis}

\address{Department of Physics, University of Crete,
P.O. Box 2208, 71003 Heraklion, Crete, Greece}

\maketitle

\pub{Received (July 30, 2002)}{Revised (*****)}

\begin{abstract}

The complex amplitudes of the electronic wavefunctions
on different sites are used as Kramers variables for describing 
Electron Transfer. The strong coupling of
the electronic charge to the many nuclei, ions,
dipoles, etc, of the environment, is modeled as a 
thermal bath better considered classically.  After 
elimination of the bath variables, the electron dynamics is described
by a discrete nonlinear Schr\"odinger equation  with norm preserving
dissipative terms  and Langevin random noises (at finite temperature).

The standard Marcus results are recovered far from the inversion point,
where atomic thermal fluctuations adiabatically induce the electron transfer.
Close to the inversion point, in the non-adiabatic regime, electron
transfer may become ultrafast (and selective) at low temperature
essentially because of the  nonlinearities, when
these are appropriately tuned.  We demonstrate and illustrate
numerically that a \textit{weak} coupling of the donor site with
an extra  appropriately tuned (catalytic) site, can trigger an ultrafast
electron transfer to the acceptor site at zero degree Kelvin,
while in the absence of this catalytic site no transfer
would occur at all (the new concept of Targeted Transfer initially developed
for discrete breathers is applied to polarons in our theory).

Among other applications, this theory should be relevant for
describing the ultrafast electron transfer observed in the photosynthetic
reaction centers of living cells.
\end{abstract}

\section{Introduction}
According to transition state theory, chemical reactions decompose into 
elementary reactions among which electron transfer is ubiquitous \cite{KU99}.
The time required for an electron transfer (ET)
between different molecules is expected to be minimum at the so-called Marcus
inversion point in the space of parameters.
However, it is precisely  in the vicinity of this inversion point that
the standard adiabatic (Born-Oppenheimer) approximation used in
Marcus theory, breaks down. We propose a non adiabatic dimer model
which should improve the Marcus theory in this regime. We also show
that our approach extended to trimer models predicts the possibility
of catalytic ultrafast electron transfer when appropriate 
conditions are fulfilled.

The Marcus theory \cite{Mar93} well describes in many cases the features
observed for ET between two molecules.
It is essentially an adiabatic theory where
the atomic fluctuations are supposed to be slow at the scale of
the characteristic time of the electron dynamics so that
the wavefunction of the electron may remain practically an eigenstate 
of the time dependent potential created by the atoms.
This adiabatic assumption is valid when
the largest phonon energy $\hbar \omega_{ph}$
remains  much smaller than the smallest excitation energy of 
the electron. In  a two level system, this energy is essentially the distance
$\hbar \omega_{el}$  between the two levels.
This condition is very often fulfilled in real systems because
electronic energies are generally much larger than phonon energies.

\begin{figure}[htbp]
    \centering
    \includegraphics[width=0.7 \textwidth]{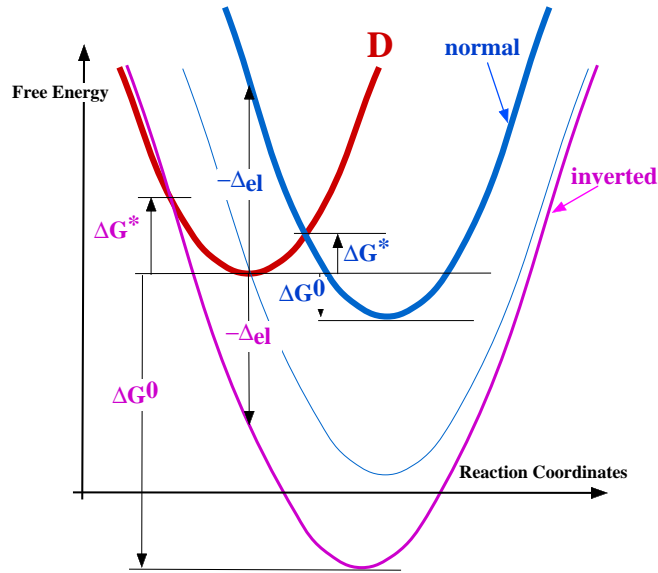}
\caption{Free energy versus Reaction Coordinates of the system Donor-Acceptor
when the electron is  on the Donor (top left curve $D$) or
on the Acceptor for several redox potentials in the normal regime (top right curve),
at the inversion point (middle right curve) and in the inverted regime
(bottom right curve).
The chemical reaction energy is the distance between the energy minima $\Delta G^0$. 
The energy barrier is $\Delta G^{\star}$. The electronic excitation energy 
on the Donor at fixed Reaction Coordinates is $\Delta_{el}$.}
\label{fig1}
\end{figure}

Since in the adiabatic case the electron dynamics is essentially driven by 
the dynamics of the atoms, ET is essentially induced by their fluctuations.
The free energy of the system as a  function of the reaction coordinates
(describing the global atomic configuration), depends on the electronic state which
could be either on the Donor molecule or the Acceptor molecule. The
well-known Marcus scheme is represented in fig.\ref{fig1} (see ref.\cite{Mar93}
for details). In the normal regime, the ET from the Donor to the Acceptor
at fixed Reaction Coordinates requires a positive energy $\hbar \omega_{el}=
-\Delta_{el}$. ET is a thermally activated process since  thermal fluctuations
of the lattice are necessary to overcome the energy barrier $\Delta G^{\star}$.
In the inverted regime, this energy  $-\Delta_{el}$ is negative. Although
ET could be achieved at low temperature by a photon emission
at frequency (energy) $\hbar \omega_{el} = \Delta_{el}$ 
(photoluminescent chemical reaction \cite{Mar93}),
activation processes above the energy barrier  $\Delta G^{\star}$
become by far more efficient and prevalent at higher temperature.

$\Delta G^{\star}$  turns out to be just zero at the inversion point when
$\Delta_{el}=0$  (see fig.\ref{fig1}). This is the regime where 
ET is expected to be at maximum speed and still effective at low temperature
because of the  absence of energy barrier. However,
in the vicinity of this inversion point
the validity of the adiabatic hypothesis necessarily breaks down,
since the characteristic energy
$|\Delta_{el}|$ of the electronic excitation becomes small.
Our approach improves the theory for ET to be valid in this 
non-adiabatic regime as well.

\section{A Non-adiabatic Model for ET}

\begin{figure}[htbp]
    \centering
    \includegraphics[width=0.7 \textwidth]{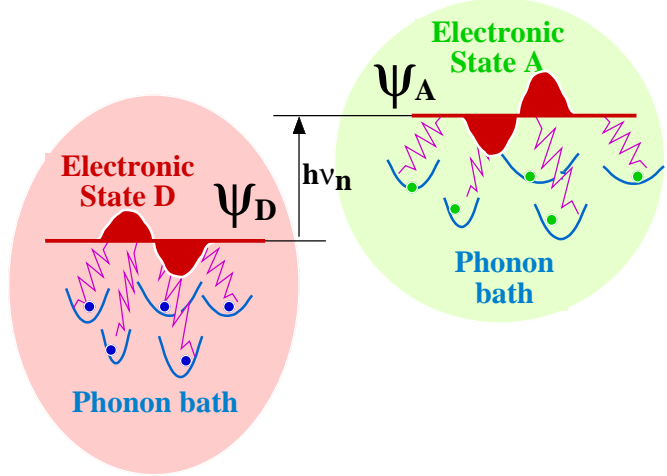}
\caption{Schematic representation of a Donor-Acceptor pair with electronic levels interacting with phonon baths
in the normal Marcus regime}
\label{fig2}
\end{figure}

We consider a single electron tunneling between  Donor ($D$)  and Acceptor
($A$) systems representing large molecules with
many vibrational degrees of freedom (see fig.\ref{fig2}). Each of these molecules
$\alpha$ ($\alpha=D$ or $A$ but more molecules may be involved) is
supposed to involve for simplicity a single electronic state with a wavefunction
$|\mathbf{\Psi}_{\alpha}> =\mathbf{\Psi}_{\alpha}(\mathbf{r};\{u_i^{\alpha}\})$,
where $\mathbf{r}$ is the space coordinate, $u_i^{\alpha}$ the phonon
coordinates, and we assume that the Born-Oppenheimer
(adiabatic) approximation. This approximation is valid when the other electronic states
on this molecule are far apart in energy from the considered state
$|\mathbf{\Psi}_{\alpha}>$ at the scale of the maximum phonon energy.
Then this electronic wavefunction can be considered as a function  of
the molecule phonon coordinates $\{u_i^{\alpha}\}$, including possibly those of
the environment and  in particular, the solvent.

Within a standard tight-binding representation, the state of the
electron in the whole system, has the form 
$\sum_{\alpha} \psi_{\alpha} |\mathbf{\Psi}_{\alpha}>$.
We use as Kramers reaction coordinates  \cite{Kra40} the complex amplitudes
$\psi_{\alpha}$ of the electronic wavefunction. We have the normalization
condition $\sum_{\alpha} |\psi_{\alpha}|^2 =1$.

It is convenient to define first the minimum  energy of the system
of the two coupled electronic states
$H_T(\{\psi_{\alpha}\})$ at fixed collective variables $\{\psi_{\alpha}\}$.
Then, the Hamiltonian of the interacting electron-phonon system can be written as 
\begin{equation}
 \mathbf{H} = H_{T}(\{\psi_{\alpha}\})  
+\sum_{\alpha}  H_{\alpha}^{ph}(\{|\psi_{\alpha}|^{2},
u_{i}^{\alpha},p_{i}^{\alpha}\})
\label{glH}
\end{equation}
The electronic density $|\psi_{\alpha}|^{2}$ on  each molecule $\alpha$ couples to the
coordinates of the same molecule assumed to be harmonic and thus consisting of a collection of
independent harmonic oscillators $i$ with position-momentum coordinates $u_{\alpha,i}
,p_{\alpha,i}$, mass $m_{\alpha,i}$, and frequencies $\omega_{\alpha,i}$
\begin{equation}
H_{\alpha}^{ph}(\{|\psi_{\alpha}|^{2},u_{\alpha,i},p_{\alpha,i}\})
= \sum_{i}\frac{1}{2} m_{\alpha,i}\omega_{\alpha,i}^2\left(u_{\alpha,i}
-k_{\alpha,i} |\psi_{\alpha}|^2\right)^2
+\frac{1}{2m_{\alpha,i}} p_{\alpha,i}^2
\label{cphbth}
\end{equation}
In principle this coupling energy  involves all possible interactions
with the atomic coordinates and in particular, the chemical energies and 
the electrostatic energies. The latter ones could be especially important in biomolecules
which are polyelectrolytes surrounded by ions and highly polarizable water.

In general  $H_T(\{\psi_{\alpha}\})$ has not the Hermitian form
$<\{\psi_{\alpha}\}|H_T|\{\psi_{\alpha}\}>$ where $H_T$ is a linear 
operator but is highly \textbf{nonlinear} as a consequence of the electric field 
and the molecule
reorganization generated by the density variations of $\psi_{\alpha}$.
It is convenient to split this Hamiltonian in several parts
\begin{equation}
H_{T}(\{\psi_{\alpha}\})=
\sum_{\alpha} H_{\alpha}(|\psi_{\alpha}|^{2})+ 
H_{f}(\{|\psi_{\alpha}|^{2}\})
+ H_{t}(\{\psi_{\alpha}\})
\label{split}
\end{equation}
where $H_{\alpha}(|\psi_{\alpha}|^{2})$ is the energy
of the isolated molecule $\alpha$ which depends  only on its electron density
$I_{\alpha}=|\psi_{\alpha}|^2$. It is sufficient to expand this energy
at second order in electronic density for obtaining the main physical features
\begin{equation}
H_{\alpha}(|\psi_{\alpha}|^{2})= \mu_{\alpha} |\psi_{\alpha}|^{2}
+ \frac{1}{2} \chi_{\alpha} |\psi_{\alpha}|^{4}
\label{locham}
\end{equation}
$\mu_{\alpha}$ is the linear electronic level at zero occupation.
$\chi_{\alpha}=\chi_{\alpha}^{C}+\chi_{\alpha}^{R}$
is the sum of two contributions. $\chi_{\alpha}^{C}$ is the positive coefficient 
for the energy of the electric field generated by the charge $I_{\alpha}$
without lattice reorganization (capacitive energy). This coefficient takes 
into account the
electronic dielectric constant $\epsilon_{\infty}$.

$\chi_{\alpha}^{R}$ is the negative coefficient of the energy gain from the
local reorganization due to the presence of the electron. This energy
involves for a part chemical bond energies which could be broken or created and
electrostatic terms involved in the static dielectric constant.
Actually, in our model (\ref{cphbth}) the reorganization energy can be 
explicitly calculated
and we get  $\chi_{\alpha}^{R}= -\sum_i m_{\alpha,i}\omega_{\alpha,i}^2
k_{\alpha,i}^{2}$.

It is essential to remark that that the sum of the two contributions 
$\chi_{\alpha}$ might be positive or negative depending whether
it is the electrostatic energies  or the chemical energies which are prevalent.
We expect for example that when the electronic state $\alpha$ belong to the 
inner shell of a transition metal ion (which could be embedded in
a large biomolecule), the electrostatic energy is prevalent so that
$\chi_{\alpha}$ is positive. When  it belongs to a chemical bond (or a ring of
bonds), it is more likely negative. Actually, only {\it ab initio} calculations could 
estimate the real values of these coefficients. The essential physical
consequence of the nonlinearity is that the electronic level on molecule $\alpha$
\begin{equation}
E_{\alpha} = \frac{\partial H_{\alpha}}{\partial I_{\alpha}}=
 \mu_{\alpha}+\chi_{\alpha}|\psi_{\alpha}|^2
\label{nlevel}
\end{equation}
depends on its electronic occupation density.

$H_{f}(\{|\psi_{\alpha}|^{2}\})$ in (\ref{split})
is a small extra term due to molecule interactions which which could be for example
the  Coulomb interaction energy
$H_{f}(\{|\psi_{\alpha}|^{2}\}) =$ \\ $ \sum_{\alpha,\beta} 
C_{\alpha,\beta} |\psi_{\alpha}|^{2} |\psi_{\beta}|^{2}$
where $C_{\alpha,\beta}$ are mutual capacitance coefficients.

$H_{t}(\{\psi_{\alpha}\})$ in (\ref{split}) is also a small extra energy term
due to orbital overlaps but it is  essential because it allows ET.
We may choose for simplicity the form
$H_{t}(\{\psi_{\alpha}\}) =$ \\ $\lambda_{\alpha,\beta}
\psi_{\alpha}^{\star} \psi_{\beta} + c.c $.
The transfer integrals  $\lambda_{\alpha,\beta}$ are assumed to be small 
(compared to differences of electronic energy levels) and may be of the order
of phonon energies.

In an ideal anti-adiabatic regime (but non-realistic) where the transfer integrals
$\lambda_{\alpha,\beta}$  between different
molecules would be much smaller than all phonon  energies 
$\hbar \omega_{\alpha,i}$,
ET would be much slower than the phonon dynamics.
Then,the phonons could be eliminated as fast variables following adiabatically
the slow electron variables
and $ H_{T}(\{\psi_{\alpha}\})$ would be the exact Hamiltonian describing
the electron dynamics through the set of nonlinear Hamilton  equations
$i \hbar \dot{\psi}_{\alpha} = \partial H_{T}/\partial \psi_{\alpha}^{\star}$.

Actually,  nonlinearities may generate energy barriers
and even in the absence of energy barriers, the absence of energy dissipation
does not allow ET to a lower energy level.
Ultrafast ET
requires an efficient energy  dissipation which can be obtained only
 by interaction of the electron with a phonon bath
in the regime intermediate between adiabatic and anti adiabatic
\footnote{The dynamical coupling of the electron with the electromagnetic field
also generates energy dissipation. However, this coupling which is weak has to be treated
quantum using Fermi golden rule which yields relatively long life time
to quantum excitations. We neglect here this energy dissipation since we are interested
in ultrafast ET.}. 

Usually, the lattice reorganization due to the presence of an electron on a molecule
is large and involves the coherent creation of many phonons. It is thus legitimate to
treat this phonon bath classical while the electron dynamics remains quantum.
The dynamical equations of the coupled system (\ref{glH},\ref{cphbth}) are
\begin{eqnarray}
i \hbar \dot{\psi}_{\alpha}&=& \frac{\partial H_T}{\partial \psi_{\alpha}^{\star}}
-\sum_i m_{\alpha,i}k_{\alpha,i} \omega_{\alpha,i}^2 (u_{\alpha,i}-
k_{\alpha,i}|\psi_{\alpha}|^2)\psi_{\alpha} \label{Scpheq}\\
\ddot{u}_{\alpha,i}&+&  \omega_{\alpha,i}^2 (u_{\alpha,i}-
k_{\alpha,i}|\psi_{\alpha}|^2)=0 \label{harmdr}
\end{eqnarray}

The harmonic motions $u_{\alpha,i}(t)$ can be explicitly 
obtained from the linear equations (\ref{harmdr}) \cite{AK02} as 
the sum of functions of
the time dependent driving force $|\psi_{\alpha}(t)|^2$ and
a solution of the equation without driving force. Actually, this
term physically corresponds to thermal fluctuations of $u_{\alpha,i}$
and thus is random. Then,
substituting $u_{\alpha,i}(t)$ in eq.(\ref{Scpheq})  yields
the fundamental equation for non-adiabatic electron dynamics
(which preserves the norm $\sum_{\alpha}|\psi_{\alpha}|^2$)

\begin{equation}
i \hbar \dot{\psi}_{\alpha} = 
    \frac{\partial H_{T}}{\partial \psi_{\alpha}^{\star}} 
+\left(\int_{-\infty}^{t} \Gamma_{\alpha}(t-\tau) \frac{d  
|\psi_{\alpha}|^{2}}{d\tau} d\tau+ \zeta_{\alpha}(t)\right) 
\psi_{\alpha}
\label{fequat}
\end{equation}
where
$\Gamma_{\alpha}(t)=\sum_i m_{\alpha,i} \omega_{\alpha,i}^2 k_{\alpha,i}^2
\cos{(\omega_{\alpha,i}t)}$.
If there are many phonon modes with a rather uniform distribution, $\Gamma_{\alpha}(t)$
can be assumed to be a smooth decaying function of time. 
It generates energy dissipation as a kernel in eq.(\ref{fequat})
(the  absorption rate  in energy 
of a  charge fluctuation at  site $\alpha$ at frequency $\omega$
is nothing but the product of the square of its amplitude with 
the Fourier transform of $ \Gamma_{\alpha}(t)$).
We also have $\Gamma_{\alpha}(0)=-\chi_{\alpha}^R$.
The time dependent  potential $\zeta_{\alpha}(t)$ is produced by
the thermal fluctuations of the lattice. It is a colored random
Langevin force with correlation function which fulfills 
$<\zeta_{\alpha}(t+\tau) \zeta_{\alpha}(\tau)>_{\tau}
= \Gamma_{\alpha}(t) ~ k_{B}T$ at temperature $T$ \cite{AK02}.

Thus, the effect of non-adiabaticity is  to 
transform the standard linear Schr\"odinger equation describing
the dynamics of the electron into 
a \textbf{nonlinear} Schr\"odinger equation (\ref{fequat}) with
 norm preserving  \textbf{energy dissipation} terms
and with  \textbf{random 
colored} time dependent  potentials generated by atomic
thermal fluctuations. 

\section{ET in the Dimer Model}

We first show that far from the Marcus inversion point we
essentially recover the basic result of the standard theory \cite{Mar93}. The initial
Hamiltonian (\ref{glH},\ref{cphbth}) restricted to a dimer model
($\alpha=D$ or $A$) readily yields the energy surfaces schematically
shown in fig.\ref{fig1}. Neglecting the small interaction energy terms between the
molecules we obtain the essential parameters of this theory which are
$ \Delta G^0 = \mu_D +\frac{1}{2}\chi_D-\mu_A-\frac{1}{2}\chi_A$,
	$\Delta_{el} = \Delta G^0 +\frac{1}{2} (\chi_D^R+\chi_A^R)=
\mu_D^{\prime}+\chi_D^R-\mu_A^{\prime}$ and
	$\Delta G^{\star} =- \Delta_{el}^2/(2(\chi_D^R+\chi_A^R))$
where $\mu_{\alpha}^{\prime}=\mu_{\alpha}+\chi_{\alpha}^C/2$.

The energy variation $E_T(I_A)=H_D(1-I_A)+H_A(I_A)-H_D(1)$ of our dimer as 
a function of the electron density $I_A=|\psi_A|^2$ on the acceptor is
\begin{equation}
E_T(I_A)=(\mu_A-\mu_D-\chi_D)I_A +\frac{1}{2} (\chi_D+\chi_A)I_A^2
\label{enerprof}
\end{equation}
and $-E_T(1)=\Delta G^0$ is the chemical reaction energy.
 \begin{figure}[h!]
    \centering
    \includegraphics[width=0.45 \textwidth]{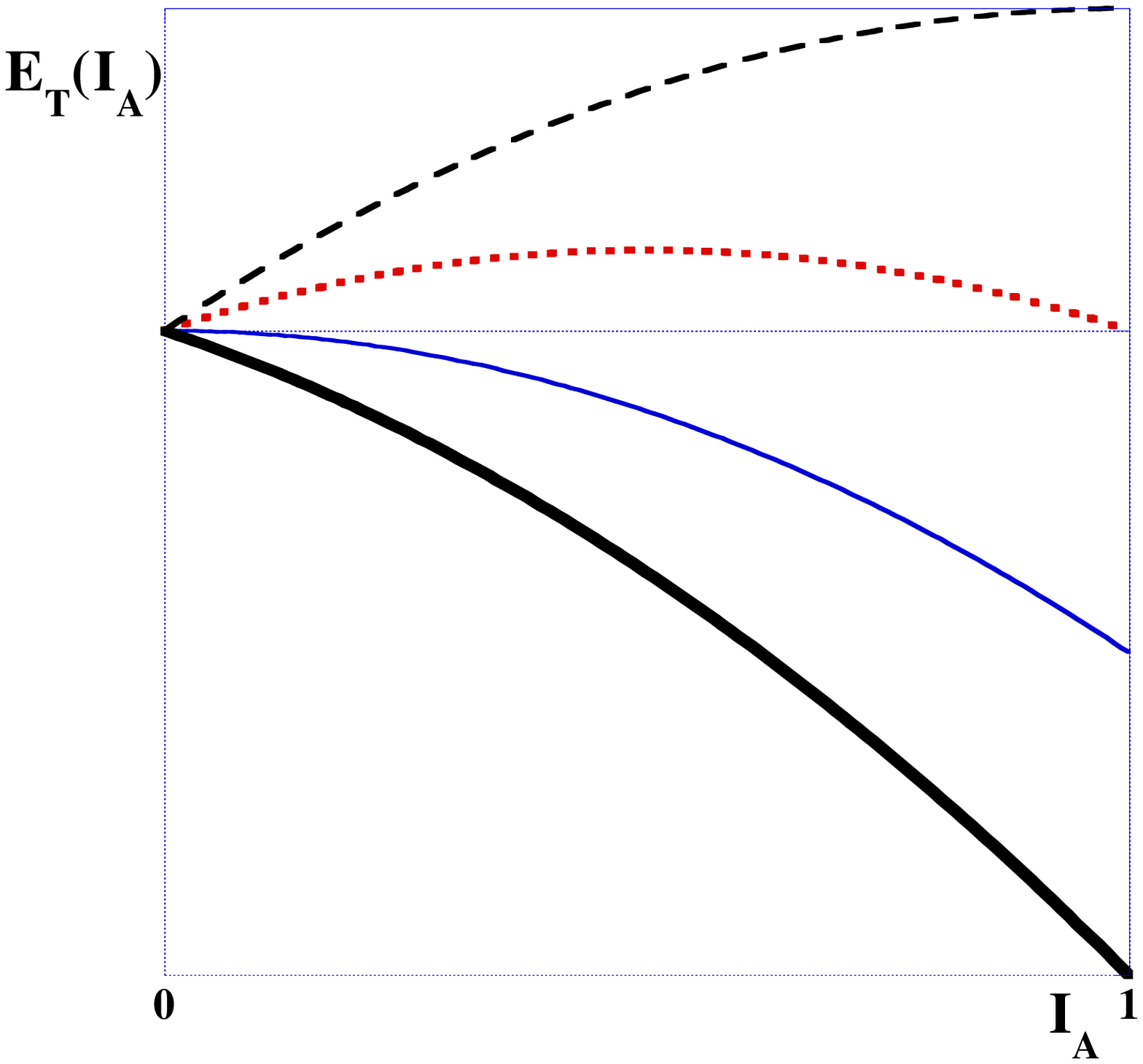}
    \includegraphics[width=0.45 \textwidth]{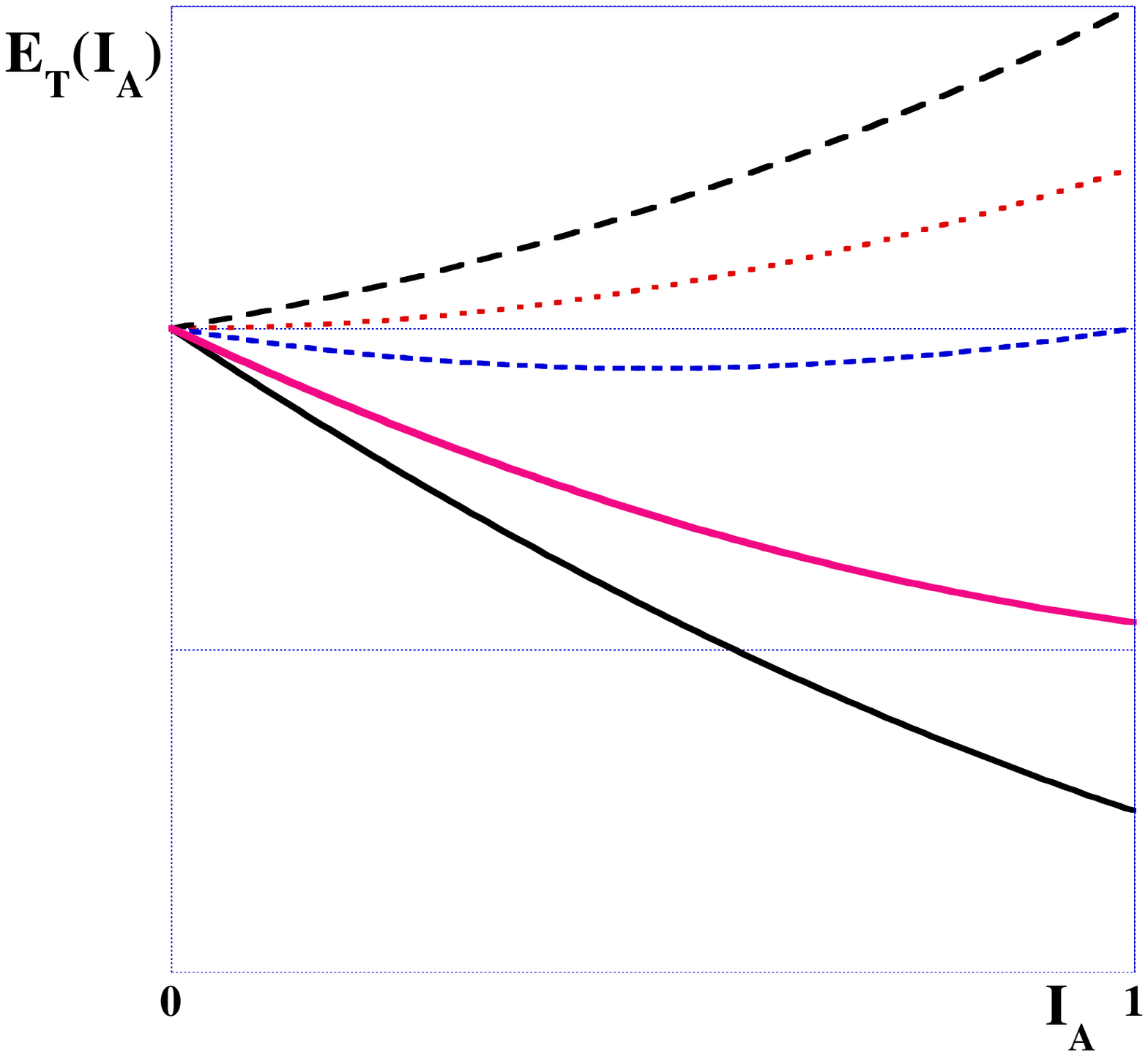}
\caption{Several energy profiles (\ref{enerprof}) of the system Donor-Acceptor
versus electron density on the Acceptor in the soft case
$\chi_D+\chi_A<0$ (left)  or in the hard case $\chi_D+\chi_A>0$ (right) }
\label{fig3}
\end{figure}
There is always an energy barrier between Donor and Acceptor 
when  $\frac{d E_T}{d I_A}(0)>0$ or equivalently 
$\mu_A<\mu_D+\chi_D$. Otherwise, the minimum of energy is not necessarily
obtained for a total transfer at the Acceptor when $\chi_D+\chi_A>0$.
The derivative  $d E_T/d I_A=E_A-E_D$ (see eq.(\ref{nlevel})) is the difference of the electronic
levels on the Acceptor and the Donor at the transfer $I_A$.
Thus, resonance between Donor and Acceptor implies a zero derivative.

For recovering the same results as  the Marcus theory from our equation, it
is essential to note that the phonon spectrum has a cut-off
at relatively small frequencies $\omega_c$. Beyond this frequency,
the Fourier spectrum of $\Gamma(t)$ is zero. Thus, when the
characteristic energy of the electron dynamics, which is the  
energy difference $E_D-E_A$ between the electronic levels, eq.(\ref{nlevel}),
is larger than the phonon energy
$\hbar \omega_c$, there is no more energy dissipation  \cite{AK02}.
Then ET cannot be achieved at zero degree K but
requires thermal fluctuations. Actually, this is the regime of validity of the
adiabatic approximation.

\begin{figure}[htbp]
    \centering
 \includegraphics[width=0.45 \textwidth]{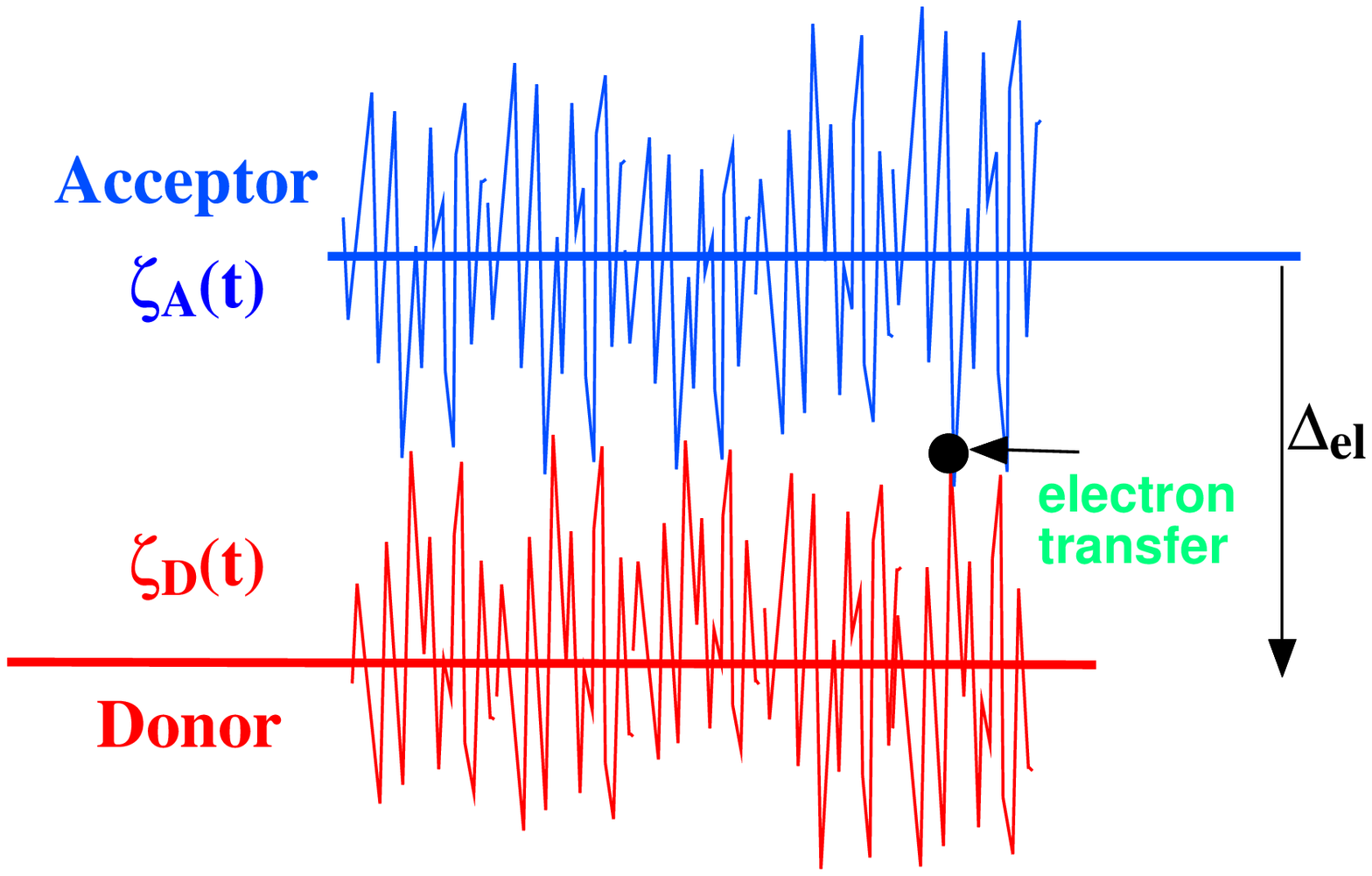}
        \includegraphics[width=0.45 \textwidth]{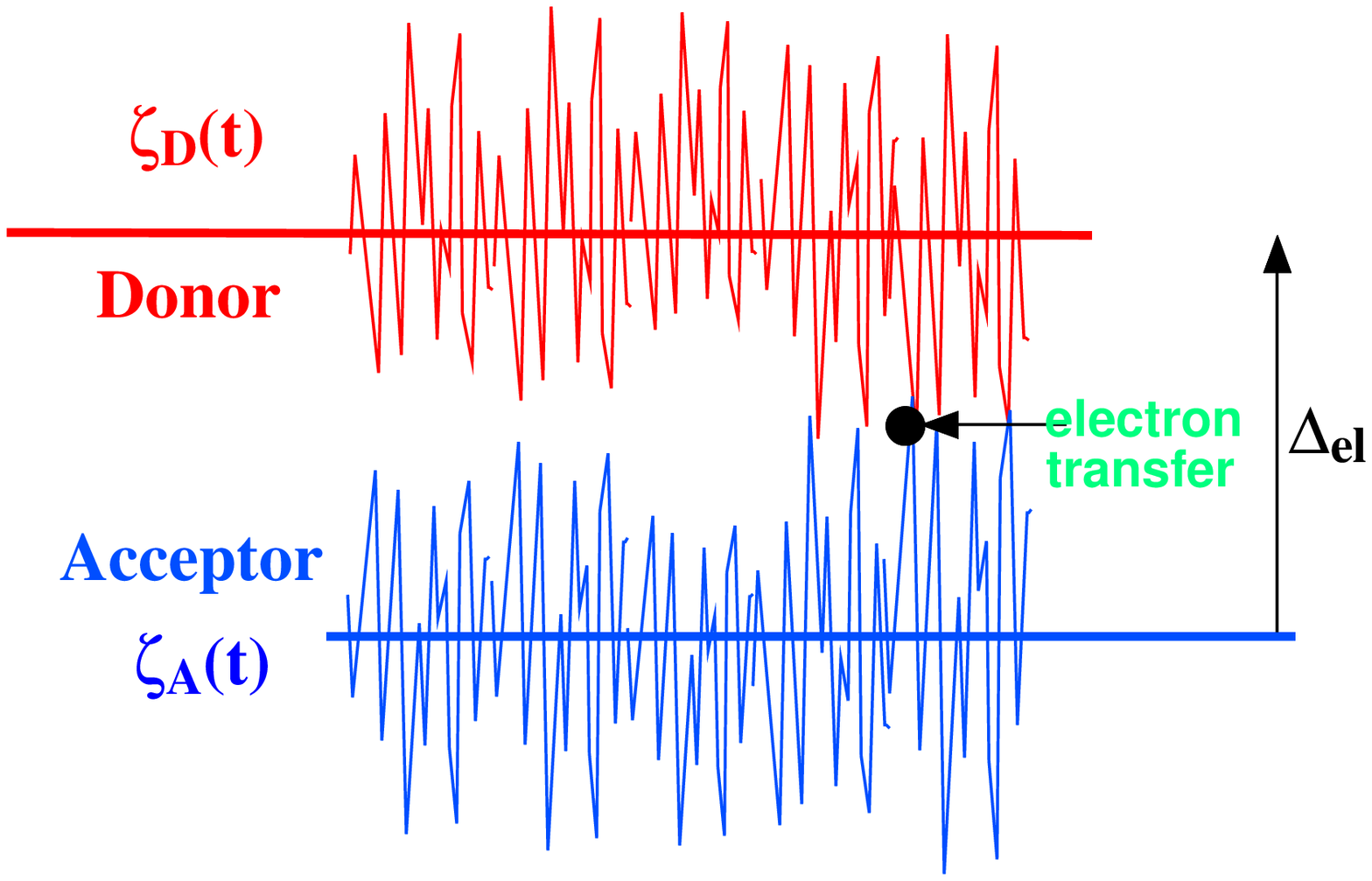}
\caption{Sketch of the electronic level fluctuations in the normal regime (left)
and in the  Marcus inverted regime(right).}
\label{fig4}
\end{figure}

The electron density which is initially on the Donor
$|\psi_D|^2=1$ and $|\psi_A|^2=0$ remains practically constant.
Eq.(\ref{Scpheq}) gives in this case the random
potential as due to phonon variables,
$\zeta_{\alpha}(t)=-\sum_i m_{\alpha,i}k_{\alpha,i} \omega_{\alpha,i}^2 (u_{\alpha,i}-
k_{\alpha,i}|\psi_{\alpha}|^2)$.
The slowly varying potential $\zeta_D(t)$ makes the  energy level of the
electron $E_D(t)=\mu_D+\chi_D+\zeta_D(t)$ on the Donor time dependent.
The unoccupied  energy level on the acceptor $E_A(t)=\mu_A+\zeta_A(t)$
fluctuates similarly.
It occurs statistically that $E_D(t) \approx E_A(t)$ or 
\begin{equation}
\mu_D+\chi_D-\mu_A+\zeta_D(t)-\zeta_A(t) \approx 0
\label{rescond1}
\end{equation}
induces an almost resonance between Donor and Acceptor so that the electron could tunnel
(see fig.\ref{fig4}).
If we discard the details concerning the probability of this tunneling process and
neglect its intrinsic time
which is generally short compared to the characteristic time for reaching  the resonance,
the characteristic  time for ET is mostly related to the time
required to reach the resonance \footnote{The electron tunneling  problem has been
considered differently in the literature which distinguishes between adiabatic processes
(strong reactants) and diabatic processes (weak reactants) \cite{KU99}.
Our non-adiabatic theory (potentially) describes both cases as well as the intermediate cases.}.

Condition \ref{rescond1} may be compared with the condition for the intersection of the two
free energy surfaces
shown in fig.\ref{fig1}, which can be written as

\begin{equation}
\mu_D^{\prime}+\chi_D^R-\mu_A^{\prime}+\zeta_D(t)-\zeta_A(t)=
\Delta_{el}+ \zeta_D(t)-\zeta_A(t)=0
\label{rescond2}
\end{equation}
This condition is similar but different from our resonance condition (\ref{rescond1}).
The reason is that the energy level variations  due to
the Coulomb energies are not  taken into account in the Marcus theory
unlike the reorganization energy
( if $\chi_{\alpha}^C=0$ conditions (\ref{rescond1}) and (\ref{rescond2}) become
identical).
Nevertheless, we can also interpret the probability of reaching the resonance as shown in fig.\ref{fig4}
\cite{AK02} in terms of an activation process with an activation energy
$\Delta^{\prime} G^{\star}=-\Delta_{el}^{\prime}/(2(\chi_D^R+\chi_A^R))
\neq \Delta G^{\star}$
where $\Delta_{el}^{\prime}=\Delta_{el}+ (\chi_D^C+\chi_A^C)/2=\mu_D+\chi_D-\mu_A$
is different from $\Delta_{el}$ because of the Coulomb terms.

\begin{figure}[htbp]
        \centering
        \includegraphics[width=0.45 \textwidth]{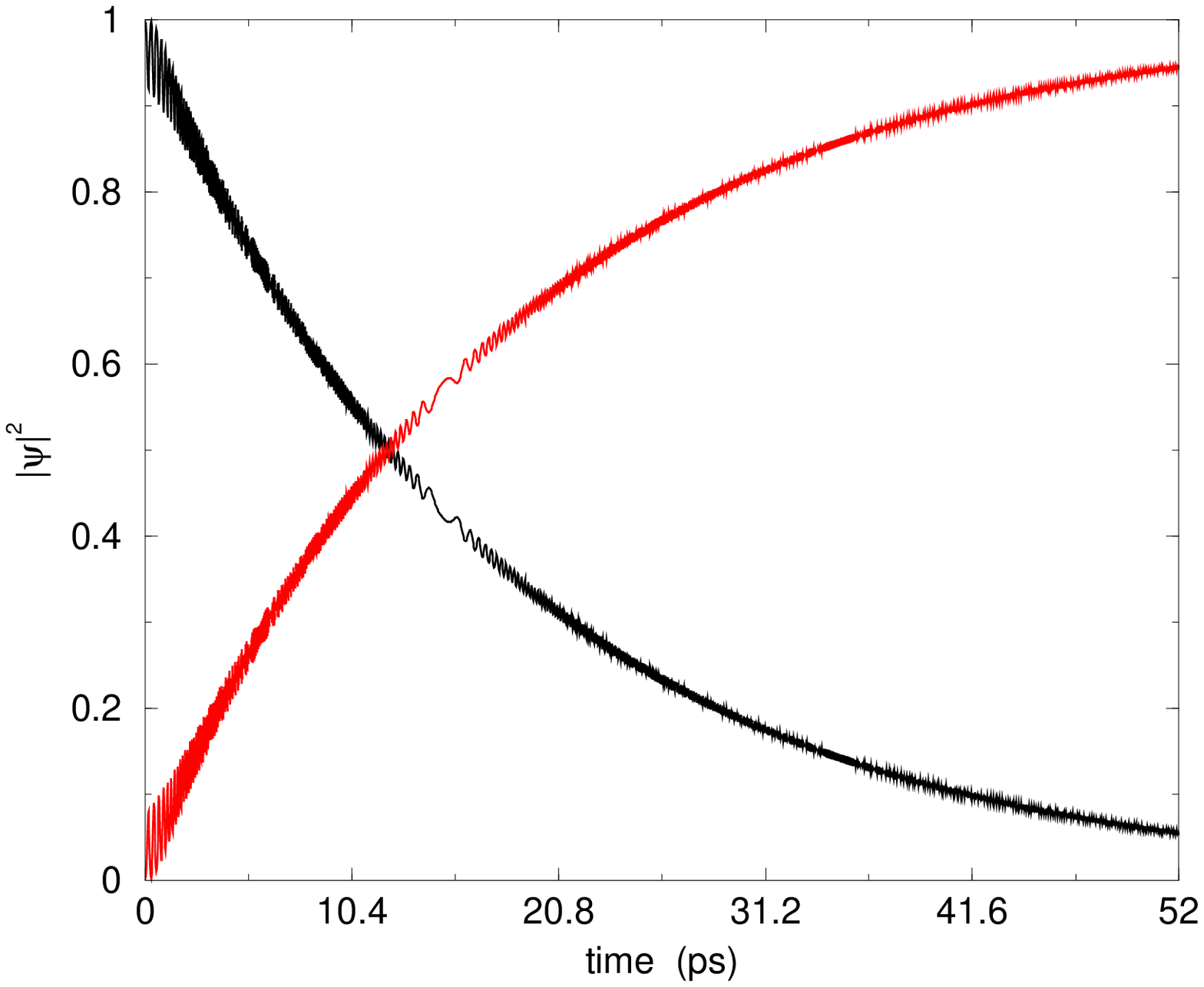}
 \includegraphics[width=0.45 \textwidth]{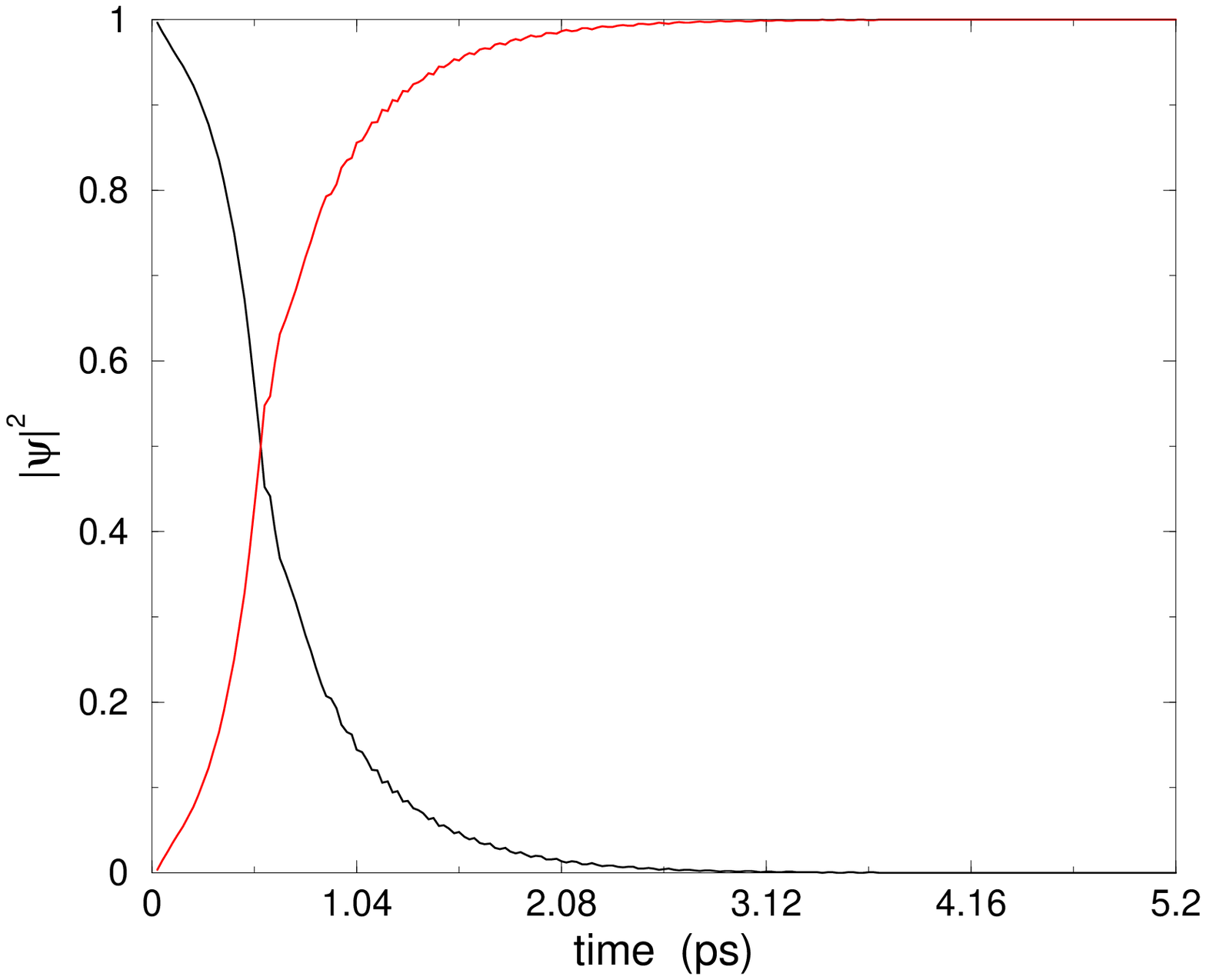}
\caption{Electron density on the Donor and the Acceptor
versus time for the dimer model at the inversion point and zero degree K where
$\mu_D=2$, $\chi_D=-1$, $\mu_A=1$, $\chi_A=-0.75$, $\lambda_{AD}=10^{-2}$,
$\gamma_D=\gamma_A=1.$ (left) or $20.$ (right)
(the time unit is 1 ps=$10^{-12}$ s for  energy in units of eV)}
\label{fig5}
\end{figure}
Nevertheless, our approach confirms the existence of an inversion point
when resonance is obtained at zero degree K for $\zeta_D=\zeta_A=0$.
Then, (\ref{rescond1}) yields $\Delta^{\prime}_{el}=0$ or $\mu_D+\chi_D=\mu_A$
which again  is different from the condition $\Delta_{el}=0$
in the original Marcus theory.

At our inversion point, the energy profile $E_T(I_A)$ has a zero derivative
at the origin $I_A=0$. Fig.\ref{fig3}
shows that there is no energy barrier only when  $\chi_D+\chi_A \leq 0$
\footnote{The only case implicitly considered in the Marcus theory
is for negative $\chi_{\alpha}=\chi_{\alpha}^R<0$.}.
We check that the initial resonance $E_D=E_A$ triggers ET.
If the derivative
$dE_T/dI_A$ remains smaller than $\hbar \omega_c$ which is equivalent to a small reaction
energy $\Delta G^0 < \hbar \omega_c/2$, ET can be achieved at zero degree K
without thermal fluctuations. Fig.\ref{fig5} shows two examples. The speed 
of ET
strongly depends on the damping which is related to the coupling to the phonon bath.
In these examples and in the following we chose for simplicity
$H_{f}(\{|\psi_{\alpha}|^{2}\})=0$ in (\ref{split}).
If the  phonon frequency cutoff  $\omega_c$ is large compared to the characteristic electronic frequencies,
a reasonable approximation is to assume that $ \Gamma(t) =2 \gamma_{\alpha} \delta(t)$ is a Dirac function. 
Then, \\
$\int_{-\infty}^t \Gamma(t-\tau)
\frac{d|\psi_{\alpha}|^2}{d\tau}  d\tau \approx \gamma_{\alpha}  \frac{d|\psi_{\alpha}|^2}{dt}(t)$ in 
eq.(\ref{fequat}). There is an optimal damping constant ($\gamma_{\alpha} \approx 40$) 
where the characteristic time required for  ET is minimum.

ET is triggered at zero degree (but slows down) when escaping only on one side
of the inversion point when $\mu_A <  \mu_D+\chi_D$. On the other side, it is blocked
at zero degree K because of the appearance of an energy barrier. However it is complete only when 
$dE_T/dI_A$ remains always negative with a modulus which  never exceeds the
phonon cutoff energy $\hbar \omega_c$ (for having efficient energy dissipation).
In  summary, our approach yields results which are qualitatively 
similar to those of the Marcus theory far from the
inversion point  but with a redefinition of characteristic parameters.
It yields more detailed features not predicted by the original Marcus
theory, close to the inversion point.

\section{Principle of Catalytic ET in a trimer model}

ET could be fast for a Donor-Acceptor system
only in special conditions close to the Marcus inversion point
and when the chemical reaction energy is small compared to the phonon 
energy cutoff.

We now show that we can take advantage of a third catalytic site 
weakly coupled to the Donor for triggering at zero temperature
an ultrafast ET  from the Donor to the
Acceptor while in the absence of catalyst, a large energy barrier
would prevent any transfer at zero degrees.

It is clear that ET could become fast only in case
of resonance or almost resonance within the phonon energy range.
Otherwise, the electronic level $E_{\alpha}$ on a molecule  depends
on its  occupation density
$E_{\alpha}= \partial 
H_{\alpha}(I_{\alpha})/\partial I_{\alpha}= \mu_{\alpha}+\chi_{\alpha} I_{\alpha}$

How to get complete ultrafast ET between
Donor and Acceptor at zero degree K not at the
Marcus inversion point? We suggest to  take advantage of the
nonlinearities for inducing relatively slow oscillations of the electronic 
level on the Donor (or on the acceptor)  which could produce a resonance
between Donor and Acceptor.

The simplest way is to
obtain these energy oscillations by thermal fluctuations as shown
fig.\ref{fig4}. This is the standard situation described above where
Marcus theory is recovered.

Another way is to induce artificially these electronic level oscillations
by exciting a specific phonon well coupled to the electronic level.
Note that this is what happens systematically when creating an exciton
by the absorption of a photon before the Franck-Condon relaxation.
This situation very likely occurs in some real systems but we
shall not discuss it here.

\begin{figure}[htbp]
        \centering
        \includegraphics[width=0.8 \textwidth]{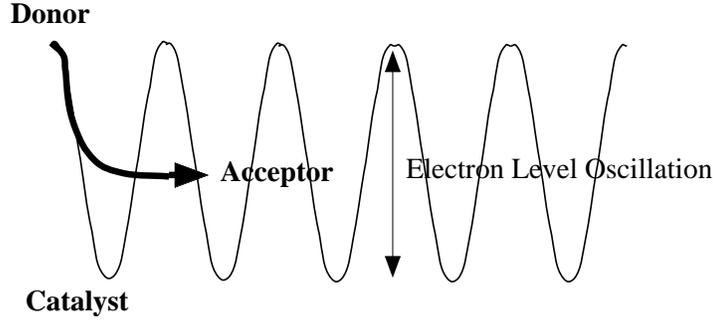}
\caption{Sketch of electronic level oscillations on the donor system inducing resonance
with the acceptor level.}
\label{fig6}
\end{figure}

\begin{figure}[htbp]
        \centering
        \includegraphics[width=0.45 \textwidth]{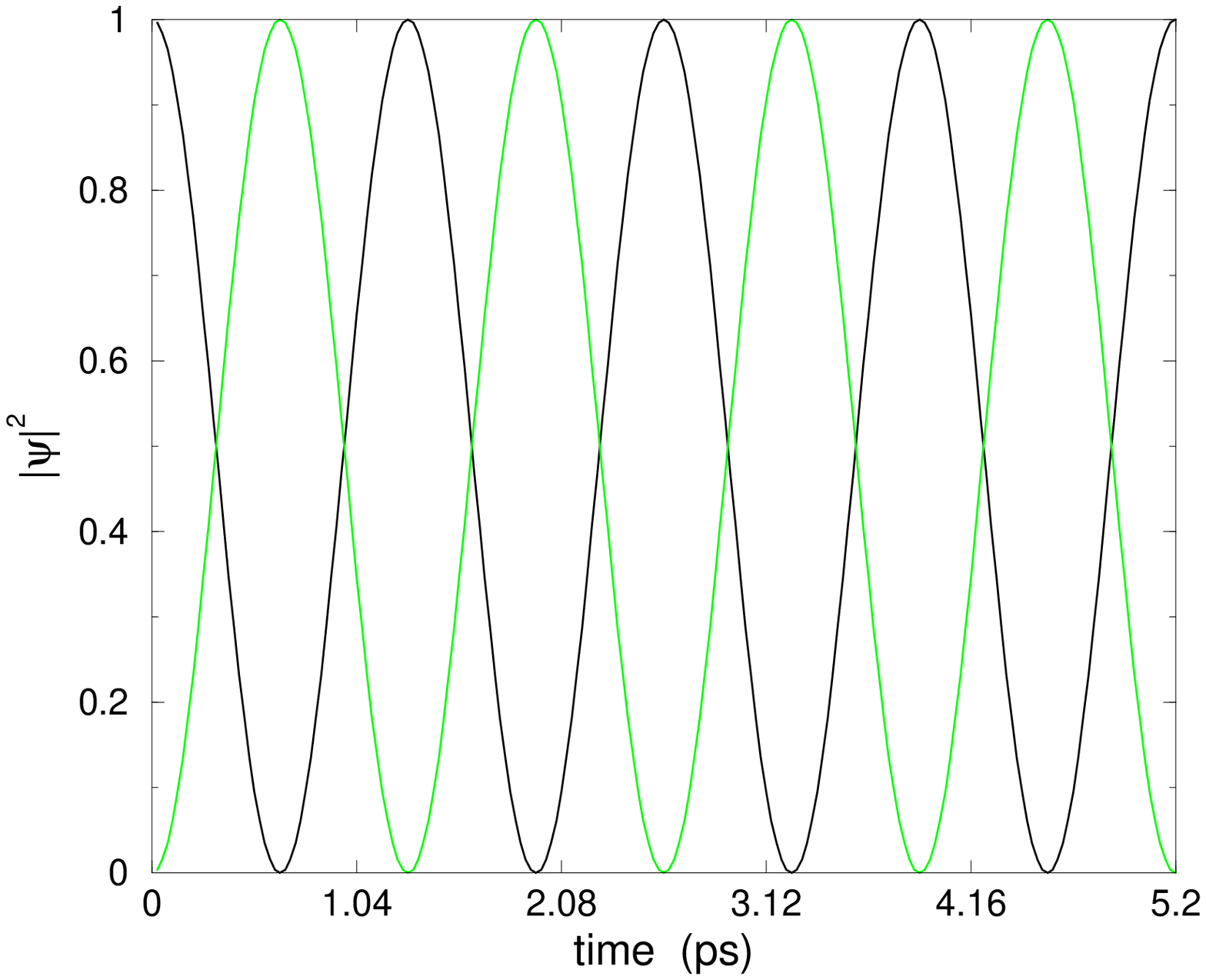}
 \includegraphics[width=0.45 \textwidth]{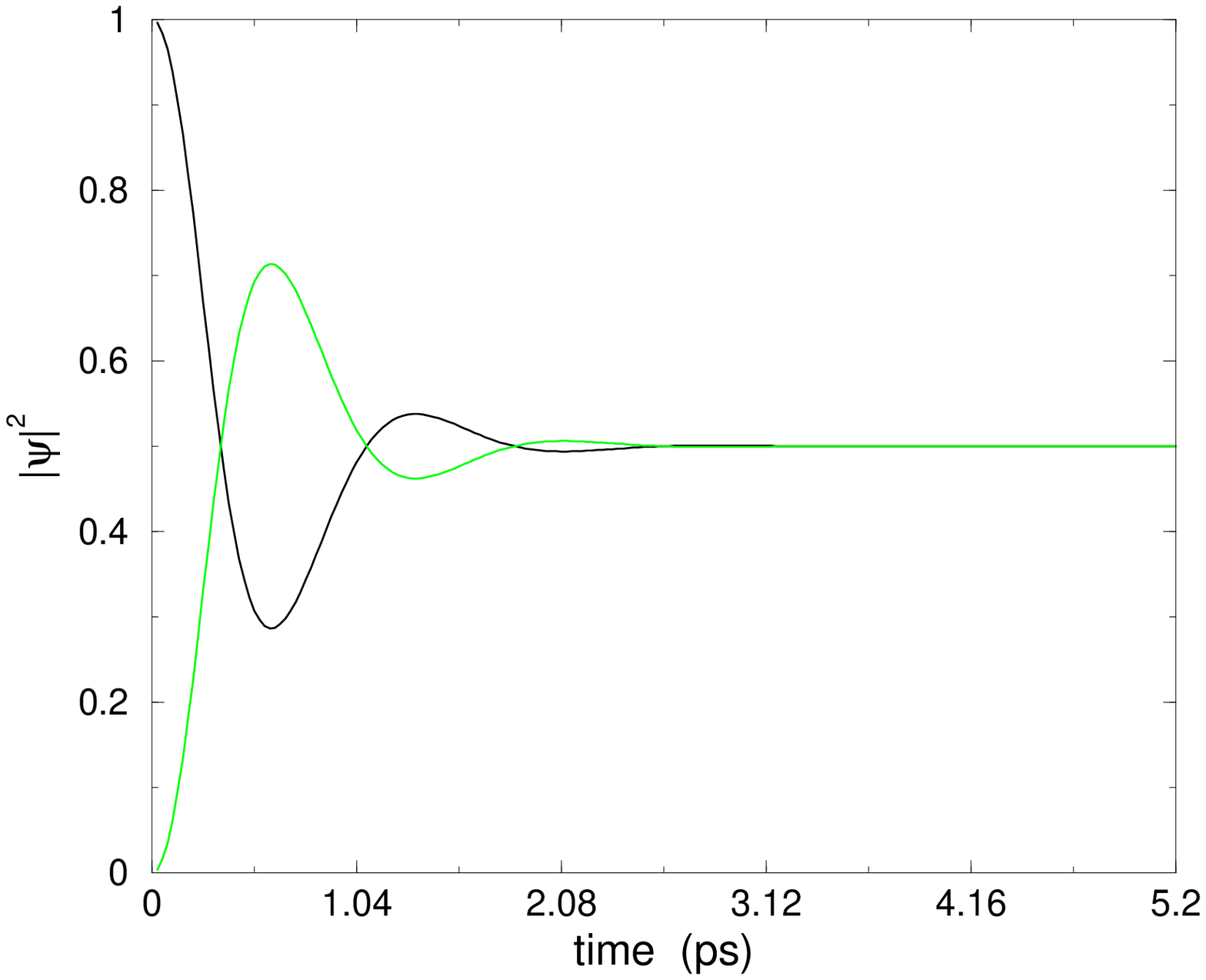}
\caption{Electron density oscillations on the Donor and its Catalyst
at zero degree K,
without damping $\gamma_D=\gamma_C=0$ (left) and with damping
$\gamma_D=\gamma_C=1$ (right). $\mu_{D}=2$, $\chi_{D}=-1$, $\mu_{C}=1$,
     $\chi_{C}=-1$, $\lambda_{CD}=10^{-2}$.}
\label{fig7}
\end{figure}

For inducing these level oscillations, we propose to use a third site,
a ``Catalyst",
which is appropriately tuned on the Donor.
For that purpose, we use the phenomenon of Targeted Energy Transfer
(or Targeted ET in this case) \cite{AKMT01,AKT01}. 
A Donor and an Acceptor at the inversion point
are in resonance but in general they 
do not  remain in resonance during the transfer. However,
there is a specially interesting  case,
 when $\chi_D+\chi_A=0$ (which is exactly solvable \cite{AKMT01})
for which the resonance persists all along the transfer.
Since $dE_T(I_A)/dI_A \equiv 0$, $E_T(I_A) \equiv 0$, the reaction energy is zero. 
As shown in \cite{AKMT01} in the absence of damping, the electron slowly
 oscillates between the Donor
and the Acceptor, which is now the  Catalyst ($C$), with a frequency corresponding to the
transfer integral $\lambda_{CD}$. The electronic level oscillates with the half period.
This situation requires to associate a soft electronic level with an
appropriate and well defined hard electronic level ( which could involve 
a metallic ion as suggested above)  ($\mu_C=\mu_D+\chi_D$ and 
$\chi_C=-\chi_D$).
When there is energy dissipation, this oscillation is damped and converges to the covalent 
state with equal density on Donor and Catalyst (see fig.\ref{fig7}). Indeed, the 
small transfer integral raises the degeneracy at zero coupling.
The range of variation of the  energy level, which is the interval
$[\mu_D+\chi_D,\mu_D]$ in the undamped case, is reduced to the interval
$[\mu_D+\chi_D/2,\mu_D]$ in the overdamped case.
However, this binding energy is very weak and negligible since
$\lambda_{CD}$ is small. Then, small thermal fluctuations become sufficient to generate
giant charge fluctuations. In practice,  Donor and Catalyst do not bind chemically but
they could trigger  ultrafast ET to another molecule.

We now test the principle of catalysis suggested by
fig.\ref{fig6}. We consider a trimer model
with Hamiltonian $$H_{tr}= \mu_D |\psi_D|^2+\frac{1}{2}\chi_D |\psi_D|^4+
 \mu_C |\psi_C|^2+\frac{1}{2}\chi_C |\psi_C|^4+
\mu_A |\psi_A|^2+\frac{1}{2}\chi_A |\psi_A|^4$$
$$+\lambda_{CD}(\psi_D^{\star}\psi_C+cc)+
\lambda_{CA}(\psi_A^{\star}\psi_C+cc)+
\lambda_{AD}(\psi_D^{\star}\psi_A+cc)$$
We assume that  Donor and Acceptor are both soft
($\chi_D<0$,$\chi_A<0$). The transfer energy is positive
which means $\mu_A+\chi_A/2<\mu_D+\chi_D/2$.

In the absence of Catalyst ($\lambda_{CD}= \lambda_{CA}=0 $)
there is a large Kramers energy barrier between Donor and Acceptor,
which implies $\mu_D+\chi_D<\mu_A$. Thus,
we are in the normal Marcus regime. ET is
impossible at zero degree K and remains slow up to relatively
high temperature.

\begin{figure}[htbp]
        \centering
 \includegraphics[width=0.45 \textwidth]{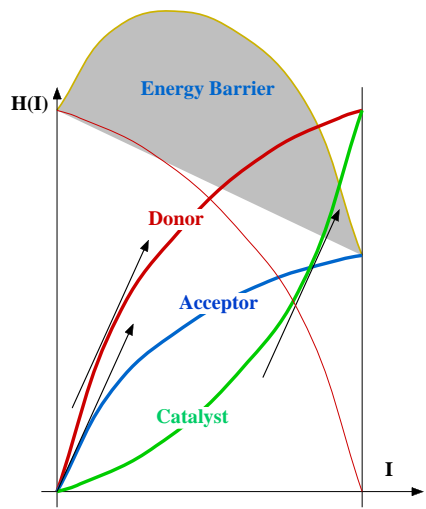}
\caption{Energies $H_D(I)$, $H_C(I)$ and   $H_A(I)$ versus electron density $I$
for the Donor, Catalyst and Acceptor in the situation of Ultrafast
ET. The energy barrier between Donor and Acceptor
without Catalyst is plotted in gray.}
\label{fig8}
\end{figure}

We now introduce the Catalyst.
 Donor and Catalyst are tuned for Targeted Electron
     Transfer which requires $\chi_{C}=-\chi_{D}$ and
$\mu_{C}=\mu_{D}+\chi_{D}$.
The initial electronic level $\mu_A$ of the acceptor
should belong to the variation interval of the electronic 
level  of the Donor-Catalyst system which yields
$\mu_D+\chi_D<\mu_A<\mu_D$ at weak damping or
$\mu_D+\chi_D/2<\mu_A<\mu_D$ at strong damping. 
Fig.\ref{fig8} sketches $H_D(I_D)$, $H_C(I_C)$ and  $H_A(I_A)$
when all these conditions are fulfilled.

\begin{figure}[htbp]
        \centering
        \includegraphics[width=0.45 \textwidth]{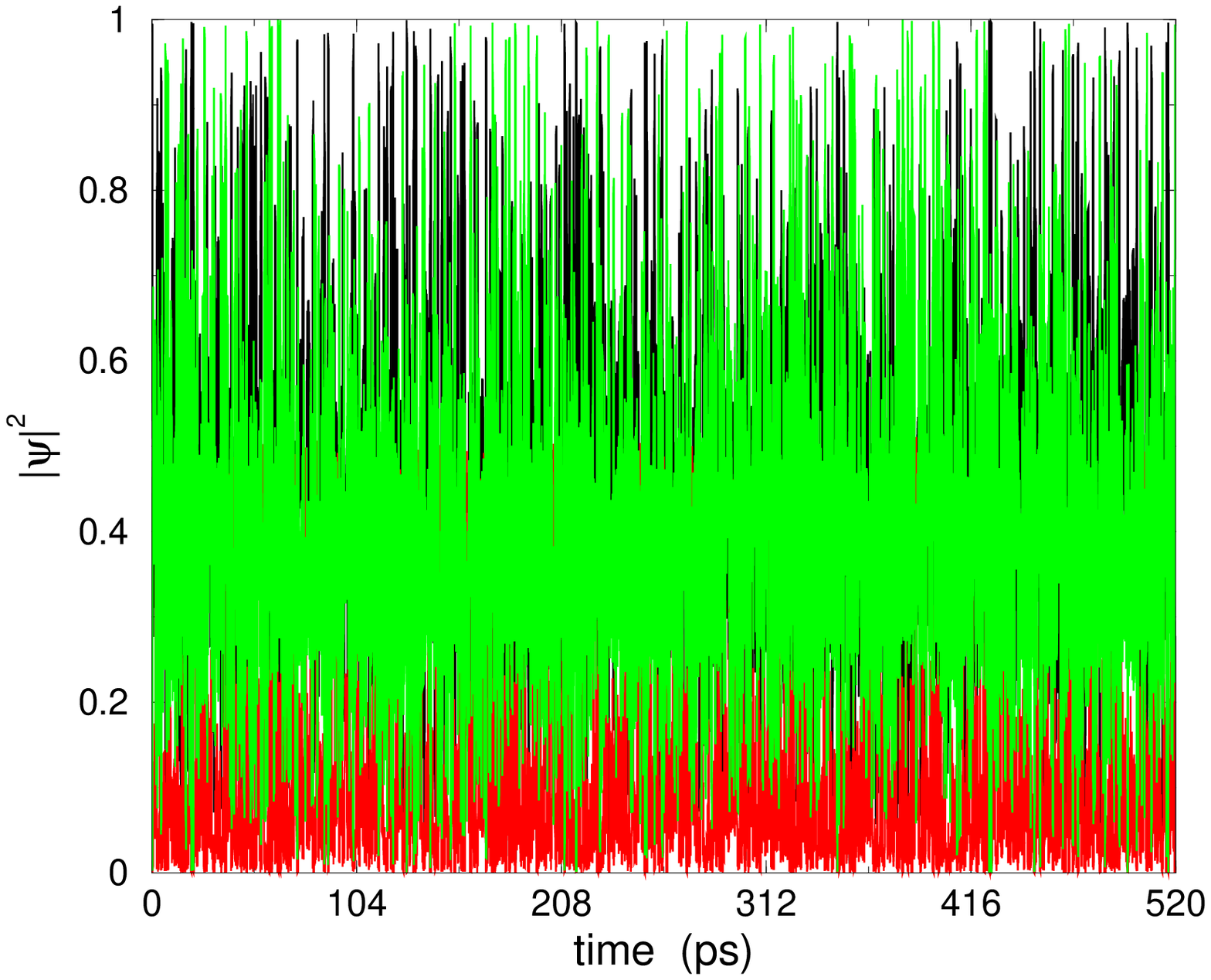}
\includegraphics[width=0.45 \textwidth]{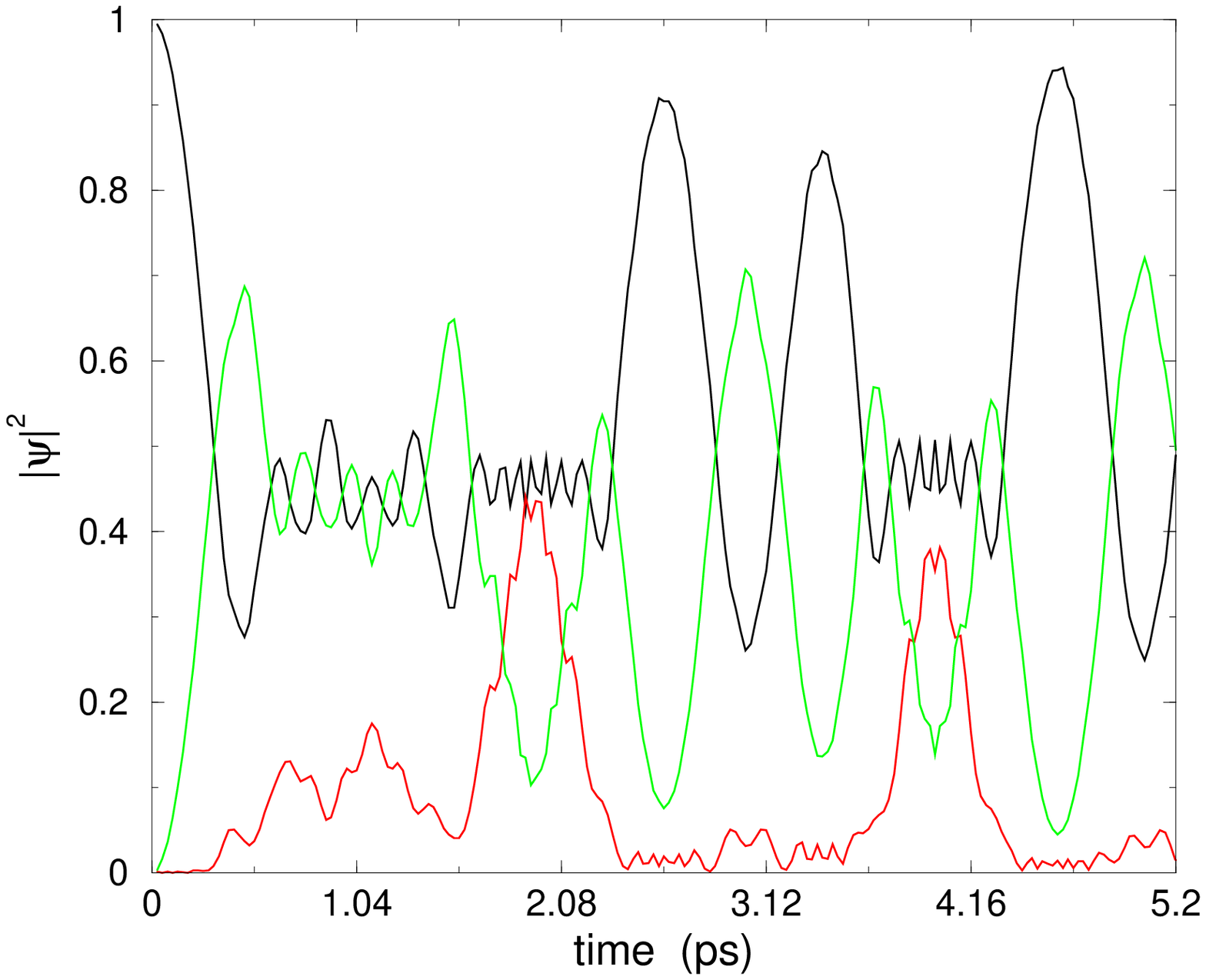}
\caption{Electron density on Donor Acceptor and Catalyst versus time
in the trimer in the absence 
of damping $\gamma_D=\gamma_C=\gamma_A=0$
at two different time scales
$\mu_{D}=2$, $\chi_{D}=-1$, $\mu_{C}=1$, $\chi_{C}=1$, $\mu_{A}=1.5$,
        $\chi_{A}=-0.75$,
 $\lambda_{AD}=\lambda_{AC}=\lambda_{CD}=10^{-2}$. The electron is initially
on the Donor.}
\label{fig9}
\end{figure}

\begin{figure}[htbp]
        \centering
        \includegraphics[width=0.45 \textwidth]{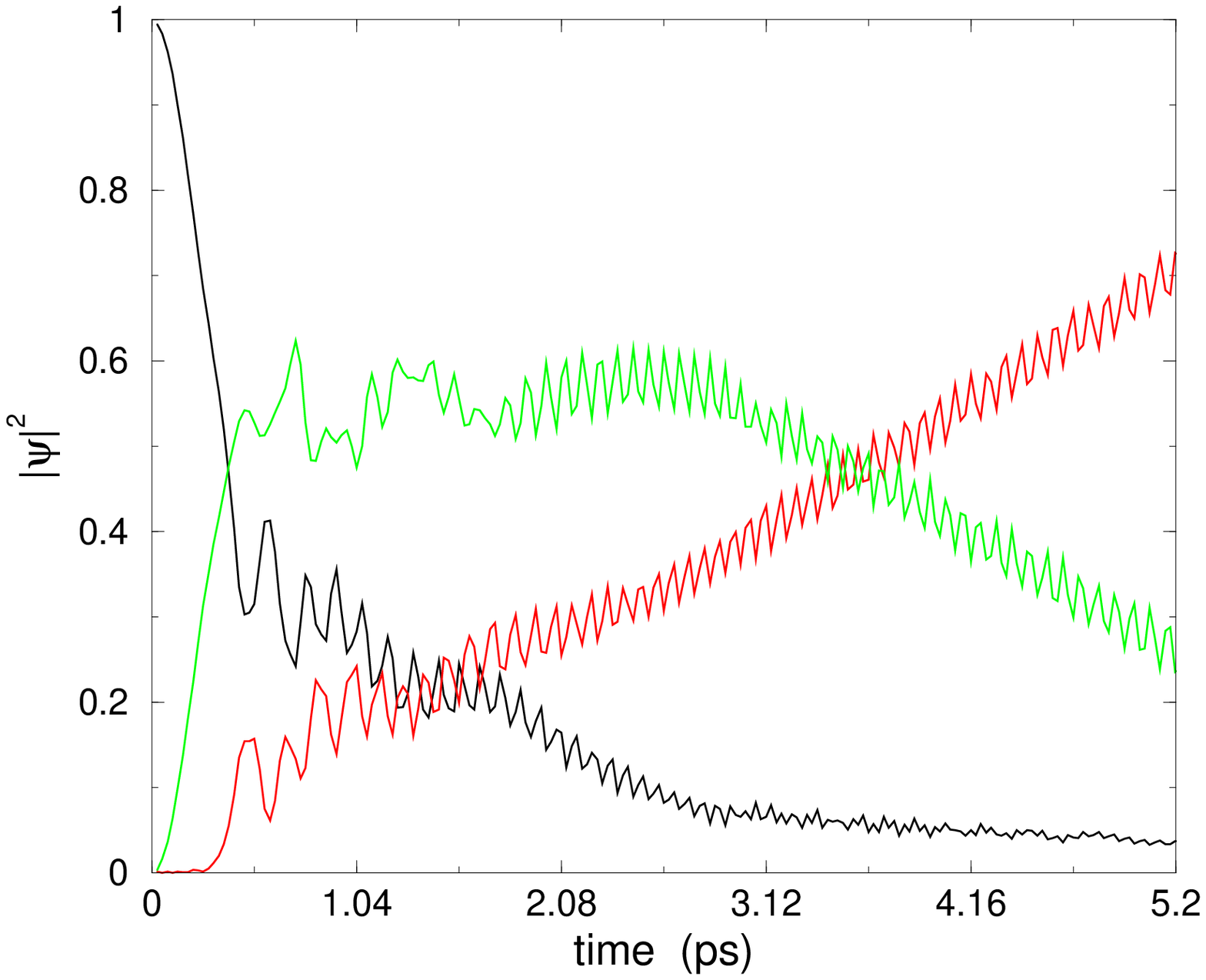}
 \includegraphics[width=0.45 \textwidth]{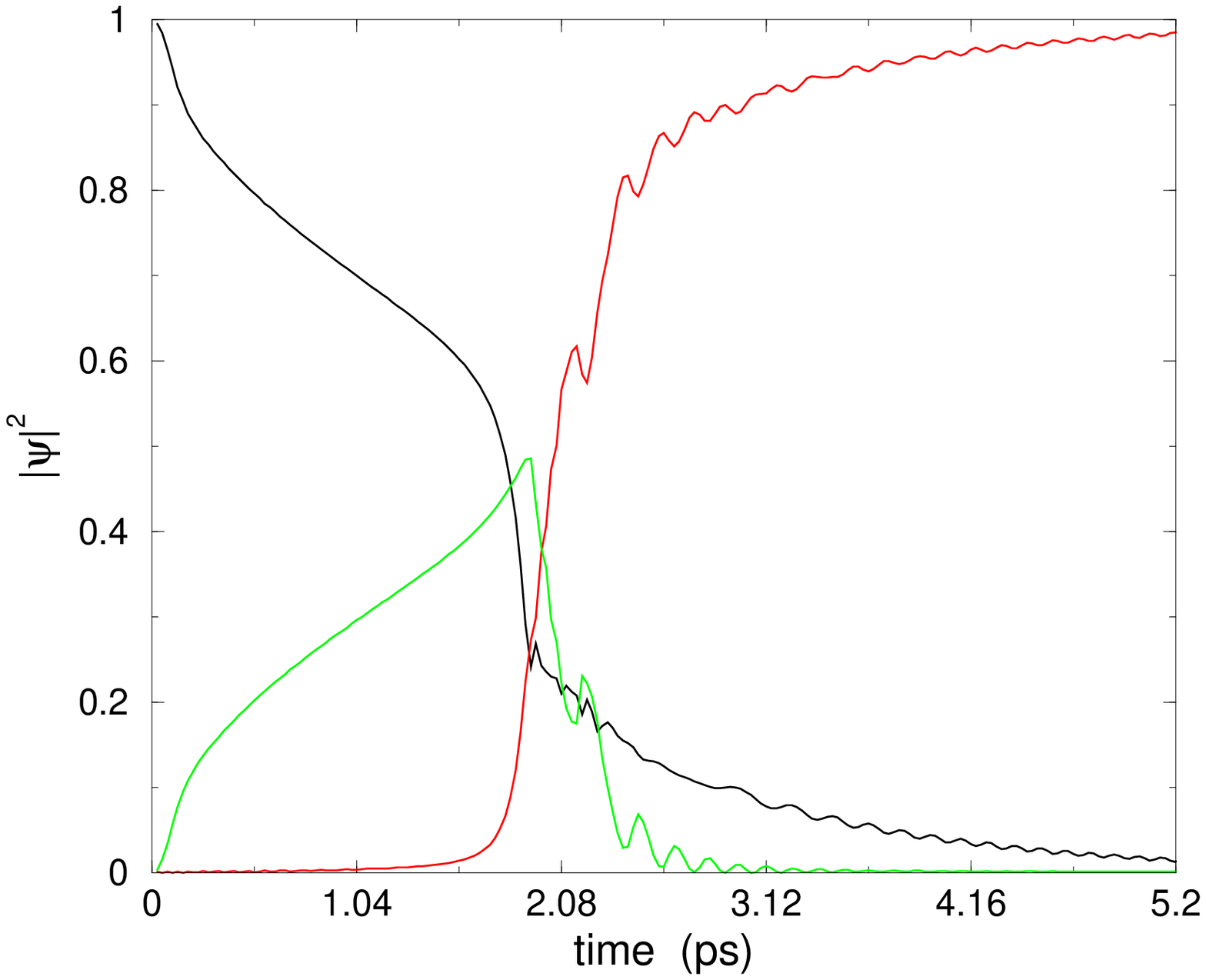}
\caption{Same as fig.\ref{fig9} but with damping
$\gamma_D=\gamma_C=\gamma_A=2$ (left) or $10$ (right)}
\label{fig10}
\end{figure}

Fig.\ref{fig9} shows that under these conditions,
huge charge fluctuations suddenly appear between between weakly coupled
Donor Acceptor and Catalyst, while 
the Donor-Acceptor system alone does not exhibit any fluctuations.
However, the absence of energy dissipation prevents the electron from falling on 
its ground-state, which is on the acceptor.
The same model with damping shows that the electron finally falls on the acceptor
while the catalyst has only taken transitively a fraction of the electronic charge (see \ref{fig10}).

This ET is highly sensitive to small perturbations of the Donor-catalyst
system which easily breaks the Targeted Transfer \cite{AKMT01,AKT01}. We have shown for example
that relatively small electric fields are sufficient for blocking ET
at zero degree K \cite{AK02}. These principles may be extended to many-site networks of electronic levels
where the electron can choose a specific path very selectively. This path
can be blocked and  switched
under small perturbations. Logical functions with one or few electrons
could be built at the molecular level suggesting potential nanodevice
applications and complex biological functions
to be studied in living cells. These studies are left for further developments.

\section{Concluding Remarks}

We presented in this short paper basic principles for 
a non-adiabatic theory of ET. We briefly sketched
new perspectives for understanding ultrafast ET
and catalysis. A more complete description of this work
with mathematical details, developments and applications
shall be published elsewhere \cite{AK02}.

A precise example of application of our trimer model
to real and well studied experimental problems
concerns the photochemical reaction in the reaction center,
where the electronic sites involved have been well identified.
In this system the photons collected in an antenna of pigment molecules
funnel to a specific site of the reaction center. Then, an
electron is ejected and transferred within few ps over
relatively long distances of about one nm (e.g. see ref.\cite{Bla02} for a review).
Further subsequent ETs follow in the biomachinery.
Besides theoretical studies and numerical
investigations of the molecules involved confirming the high efficiency of this
molecular system, the Marcus theory is not sufficient
for a global understanding of all experimental features \cite{Mar93}.
We shall propose that, out of the apparent system complexity, our simple basic 
principles could help understand
ultrafast ET as well as the puzzling features associated 
with it \cite{AK02}.

\end{document}